# Tangled in entanglements

Arto Annila[1,2,3,*] and Tarja Kallio-Tamminen[4,5]

[1]Department of Physics, [2]Institute of Biotechnology and [3]Department of Biosciences, FI-00014 University of Helsinki, Finland
[4]Physics Foundations Society and [5]Society for Natural Philosophy, Finland

**Abstract**
Two well-known conceptual conundrums of quantum mechanics referred to as instantaneous action-at-a-distance and inseparable wave-particle character are tackled using the principle of least action. Since any measurement is an action, it is reasoned that the spin of a particle just as the polarization of a photon remain indetermined for the observer until at least a quantum of action flows from the object to the observer. The detection places the quantity in question in the observer's frame of reference. This reference frame for one photon will instantaneously apply also to the polarization of the other photon provided that the two photons have not been perturbed ever since the correlated pair emerged from a radiative decay. The wave-particle duality of a single photon or an electron, that the double-slit experiment demonstrates, can also be understood in terms of actions. The energy density difference between the source of particles and their sink at the detector is a driving force that will cause flows of energy densities such as photons or electrons to propagate down along the paths of least action. Since no space is empty of energy density, the propagating particle as a flow of density will invariably perturb surrounding densities. When the driving density difference including the perturbation is leveling off via two or more paths, the ensuing flows of densities depend on each other. Therefore the flows through the two slits can never be separated from each other which will manifest as the inseparable particle-wave character.

Spooky action-at-a-distance, one of the weirdest quantum mechanical quandaries, has been chewed up so thoroughly that it shows a bad taste even to attempt to take a bite of it – or eventually a complete ignorance on the nature of the body in question. Since many old stagers have already blunted their nails when trying to get a good hold on the correlation without a force carrier, let us imagine how the puzzling phenomenon[1,2] would appear to a rookie knowing only basic physics. We would specifically like to understand, why the measurement of a photon's polarization will instantaneously reveal also the polarization of its mate photon that has emerged from the same radiative decay.

**Phases exist relative to references**
Since any observation requires some flow of energy, an observer must capture at least one photon from an object. Alternatively the observer may perceive the object by contributing at least one photon to it. Thus our Rookie reasons that the measurement as an energy transduction process will move the object from an initial state either to a final state that is down in energy or to another final state that is up in energy – which one depends on whether the observer is in the absorptive or emissive phase relative to the object. Therefore Rookie finds no need for the prevailing presumption that the unperturbed object would be in some superposition of states. The indeterminacy in the outcome is not contained in the initial state of the object as such, but follows from randomness in the phase between the object and the observer which in turn directs the flow of energy along a path that ends to a final state among alternatives.

When the measurement is understood as an energy transduction process, it follows that the polarization of a photon is established only relative to a frame of observation (Fig. 1). Thus the spin indeed remains indetermined, as stated by the EPR paradox, until the detection relates one or the other of the two correlated particles to the observer's frame. The indeterminacy with respect to the observer will prevail, and the particular polarization of one photon relative to the other photon will survive as long as the observation will perturb neither one of the two photons that resulted from the same decay. In the absence of forces the phase ($\varphi = \pi$) between the two photons remains constant in accordance with the Newton's law of constant motion ($d_t\varphi = 0$). The entanglement, i.e., the correlation between the two particles is fragile. When the two particles experience unequal energy density gradients, the degree of order will invariably decrease. In view of that, to preserve the crucial correlation does not require the presence of a force carrier which in fact would destroy it.

Thus Rookie indeed concludes that no flow of energy, i.e., a causal connection, is required to deduce instantaneously the spin of the other particle from the one referenced at the measured site provided that their mutual orientation has remained the same ever since the two emerged as a correlated pair[3,4]. However to ensure that the detected photons in fact were a correlated pair, communication from one site of a recording to the other site of detection is needed. The mandatory message can at best flow at the speed of light since Rookie finds it difficult to imagine any information without some form of physical representation[5,6]. Hence a flow of information as a flow of energy will change the receiver state from that where recordings of a photon polarization are random to that where correlations are recognized on the basis of the information given about the mate photon.

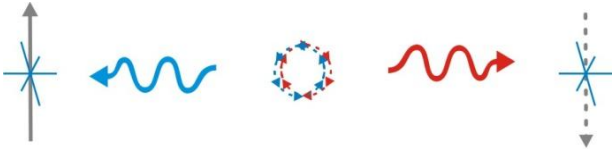

**Figure 1** A decay process (at the center) outputs two photons (blue and red) which propagate in the opposite directions. As long as neither one of the two is subject to forces the polarization of one photon will remain intact relative to its mate but indetermined relative to any external frame of reference (an axis system in blue). The detection process will place the polarization of one photon in the reference frame relative to the observer and the same frame applies at once also to the other photon provided that the correlated pair has remained unperturbed ever since it emerged from the decay.

The resolution of the EPR paradox as given above does not involve hidden variables, merely concepts of energy and time that together define an action[7] $S = \int \mathbf{p}\cdot d\mathbf{x}$ which integrates momenta $\mathbf{p}$ along paths $d\mathbf{x} = \mathbf{v}dt$ where the kinetic energy $2K = pv$ landscape is in motion with velocity $\mathbf{v}$ during time $dt$. An observation, just any other flow of energy, will drive the landscape in evolution from a stationary state toward another since at least the quantum of action, corresponding to the Planck's constant $h$, would be consumed (Fig. 1). The indeterminacy $\Delta 2K\Delta t \geq h$ is inherent in the detection because no state can be determined without a change of the state at least by $h$. A macroscopic system does not mind much of losing few quanta but a microscopic system will suffer severely, eventually going extinct when losing the very last quantum of action to the observing surroundings. This impact of detection on the object has been phrased memorably by Pascual Jordan *Observations not only disturb what is to be measured, they produce it …*[8].

**Inseparable forces and flows**
It is time to dissect another skeletal subject whereabouts hardy hyenas are surely expecting Rookie to get its teeth caught between slits. A double-slit should suffice. We would specifically like to understand, why the wave and particle natures of light and other forms of energy densities are inseparable.

Knowing that flows of energy direct to diminish energy density differences in least time, it follows that the flows of energy densities, such as a stream of photons or electrons from a high density source, will disperse along the paths of least action, for example, those passing via two slits. The flows of energy densities over a time interval $t$ will consume the driving forces due to the scalar $U$ and vector $Q$ potential differences, and the balance is maintained by a change in the kinetic energy $2K$. The conservation among the three forms of actions $2Kt = -Ut + Qt$ was conjectured a long time ago[9,10] but perhaps the balance equation is easiest to recall in a differential form where an electron is falling with velocity $\mathbf{v}$ down along electric field $-\mathbf{v}\cdot\nabla U$ and dissipating energy to the surroundings as light down along vector potential gradient $\partial_t Q$ orthogonal to the electron's directed path[3,11]. Curiously though, when energy disperses from a source down along two or more paths, the natural process will be intractable. A flow by the mere act of flowing will decrease the driving density difference and when the same gradient forces also other flows, these will be affected as well and vice versa[12]. Therefore when the flows via alternative paths are consuming the same energy density difference, they will depend on each other.

When the quantized flows of energy are leveling off the density difference, the evolutionary progress is measured by a change $d_t P = d_t \int \psi^* \psi d\mathbf{x}$ in the probability $P$. So the probability is physical. The wave function $\psi(\mathbf{x},t)$ is often regarded merely as abstraction but here it turns out to be a particularly fitting formalism, via its mutually orthogonal spatial and temporal variables, to describe a flow of energy density from one locus of energy density across $\mathbf{x}$ to another locus during time $t$. When the flows level off the density differences, the wave functions will change and eventually the natural process will attain the state where the energy landscape has no curvature[3]. Then the system has arrived at a thermodynamic steady state $d_t P = 0$ where the opposite circulations of energy densities $\psi$ and $\psi^*$, as familiar from Kirchhoff's law, are equally abundant on their common trajectory. The conserved quantities of the stationary state, most notably mass, relate to the symmetry group of the circuit via Noether's theorem[13].

It does not occur to Rookie that some form of an energy density could possibly be a point-like singularity that would fit in to its surroundings without causing any perturbation.

On the contrary, a photon has its wavelength and also the electron's finite magnetic moment and charge imply some finite-sized of a circulation whose energy density will invariably perturb surrounding energy densities. Since no space is without some energy density, at least the vacuum density will be perturbed by the flows of densities, for example in the form of light or electrons that propagate down along the available paths from a source to the sink that acts as a detector. The all-around hovering universal energy density, characterized by permittivity and permeability that define the speed of light $\varepsilon_o\mu_o = c^{-2}$ and impedance $\varepsilon_o/\mu_o = Z^2$, is familiar from Casimir effect[14] and Aharanov-Bohm experiment[15]. Thus Rookie understands that as long as the two slits are within the spread of an energy density perturbation, the flow of energy density through one slit depends on the flow through the other slit because both streams consume the common density difference. When one slit is conducting, the density difference across the other slit will also change and affect the other flow, and vice versa (Fig. 2). Therefore inseparable interference effects will arise, when the coherent flows recombine after they have taken two or more paths to maximize the overall dispersal of energy in the least time[16]. Moreover, any attempt to sniff, how the flows of energy density distribute between to the two slits, will require, just as any measurement, some flow of energy, which in turn will obscure the interference pattern by contributing incoherently to the energy dispersal process.

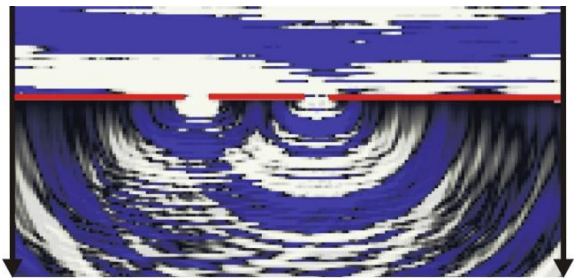

**Figure 2** Photons (or electrons) are energy densities (depicted as contrast variations in the sourrounding density) that propagate via two available paths (openings in the red wall) down along the energy density gradient between a source (up) and a sink (down). A propagating energy density will perturb the surrounding vacuum energy density and thereby also induce energy density differences that in turn act as driving forces. When the density perturbation extends from one slit over to the other, the two flows that are consuming it will depend on each other. Since the flow of energy density through one slit will affect the density difference across the other slit, the paths are interdependent and give rise to inseparable particle-wave character. (http://www.falstad.com/ripple/)

The interdependence between the flows of energy densities and the energy density differences as their driving forces is familiar also from the three-body problem and from the traveling salesman problem. When there are three or more degrees of freedom, irreversible processes are intractable because the forces and flows cannot be separated from each other for the integration to a closed form. Only when a single path is provided, i.e., when there are only two degrees of freedom, will the energy dispersal process be deterministic. Physics mostly focuses upon these special cases of bound actions known as the Hamiltonian systems where the forces can be separated from the flows and the motions can be tracked by integration.

**Actions embody space and time**
Undoubtedly there are many more challenging corpses around for our Rookie to poke into, for example those at Zeno's[17], but we have seen enough to know the Rookie's way of applying the universal law of maximal energy dispersal. Rookie does not gaze into a single piece of game but sees reality in a holistic way so that diverse systems take part in the overall process of energy dispersal by interacting with their surrounding systems[18]. A spatially localized energy density is a closed action composed of one or multiples $\hbar$. It is surrounded by other actions. Their mutual energy density differences are the driving forces that will diminish with time whose flow, in turn, is a flow of energy density in the form of open actions composed of one or multiples $h$. The open actions carry energy from high-density closed confinements to others of lower densities. These natural processes, even when the equation of motion is intractable, will naturally select from the available variation the least-time paths of dispersal, known also as geodesics[19,20]. Our Rookie, when facing the diversity of natural phenomena, is easily lost in case-by-case reasoning garnished with fancy formulas, but moves in for the kill applying the general principle of least action that may be too transparent to catch the eyes of those tangled in entanglements.

**Acknowledgement**
We thank Stanley Salthe and Juha Samela for insightful corrections.